\documentclass[usenatbib]{aa}
\usepackage{graphicx}	
\usepackage{amsmath}	
\usepackage{hyperref}
\usepackage{xcolor}
\usepackage{txfonts}

\begin{document}
\title{Where Giants Dwell: Probing the Environments of Early Massive Quiescent Galaxies}
\author{Gabriella De Lucia\inst{1,2}\fnmsep\thanks{e-mail:gabriella.delucia@inaf.it}, Lizhi Xie\inst{3}, Michaela Hirschmann\inst{4,1} \and Fabio Fontanot\inst{1,2}
}
\institute{INAF - Astronomical Observatory of Trieste, via G.B. Tiepolo 11, 
        I-34143 Trieste, Italy
        \and
        IFPU - Institute for Fundamental Physics of the Universe, via Beirut 2,
        34151, Trieste, Italy
        \and
        Tianjin Normal University, Binshuixidao 393, 300387, Tianjin, People’s
        Republic of China
        \and
        Institute for Physics, Laboratory for Galaxy Evolution, EPFL,
        Observatoire de Sauverny, Chemin Pegasi 51, 1290 Versoix, Switzerland
}

   \date{Received ???, 2025; accepted ???, 2025}

   \abstract{We investigate the environments of massive quiescent galaxies at
     $3 \lesssim z \lesssim 5$ using the GAlaxy Evolution and Assembly (GAEA)
     theoretical model. We select galaxies with $10^{10.8}\,{\rm M}_{\sun}$ and
     specific star formation rate below $0.3\times t^{-1}_{\rm Hubble}$,
     yielding in a sample of about $5,000$ galaxies within a simulated volume
     of $\sim 685$~Mpc. These galaxies have formation times that cover well the
     range inferred from recent observational data, including a few rare
     objects with very short formation time-scales and early formation
     epochs. Model high-z quiescent galaxies are $\alpha$-enhanced and exhibit
     a wide range of stellar metallicity, in broad agreement with current
     observational estimates. Massive high-z quiescent galaxies in our model
     occupy a wide range of environments, from void-like regions to dense knots
     at the intersections of filaments. Quiescent galaxies in under-dense
     regions typically reside in halos that collapsed early and grew rapidly at
     high redshift, though this trend becomes difficult to identify
     observationally due to large intrinsic scatter in star formation
     histories. The descendants of high-z massive quiescent galaxies display a
     broad distribution in mass and environment by $z=0$, reflecting the
     stochastic nature of mergers. About one-third of these systems remain
     permanently quenched, while most rejuvenation events are merger-driven and
     more common in overdense regions. Our results highlight the diversity of
     early quiescent galaxies and caution against assuming that all such
     systems trace the progenitors of present day most massive clusters.}
   \keywords{galaxies: formation -- galaxies: evolution -- galaxies:
     high-redshift -- galaxies: star formation -- galaxies: stellar content --
     galaxies: abundances}

\titlerunning{On the environment of early massive quiescent galaxies}\authorrunning{G.~De Lucia et al.}
   
   \maketitle
%
\section{Introduction}
\label{sec:intro}

The assembly and quenching of massive galaxies are central questions in modern
astrophysics. In the local Universe, where star formation can be measured
accurately for large samples of galaxies using a variety of indicators,
galaxies are found to follow a `bimodal' distribution
\citep[e.g.][]{Blanton_etal_2003,Kauffmann_etal_2003,Baldry_etal_2004}.  This
bimodality exhibits a well defined red sequence dominated by low star formation
rates, and a blue cloud populated by galaxies that are still actively forming
stars at non negligible rates.  The wealth of observational information
available in the local Universe has allowed us to characterize how the
distribution in the sSFR/color versus stellar mass plane depends on the
environment, defined either in terms of local galaxy density or through a more
`global' environmental proxy such as halo mass or large-scale over-density. The
fraction of passive galaxies is found to be larger in over-dense regions
\citep[e.g.][]{Kauffmann_etal_2004,Balogh_etal_2004,Baldry_etal_2006},
indicating that environmental processes had a significant impact on at least a
fraction of the quenched galaxy population observed today.

At earlier cosmic epochs, accurate measurements of the star formation are more
difficult to obtain for statistical samples of galaxies, and a less accurate
environmental characterization is typically possible due for example to only
photometric redshift information available or sparse spectroscopic sampling
rates. Yet, available data lead to general trends that are similar
to those observed in the local Universe, with an overall decrease of the
fraction of quiescent galaxies at earlier cosmic epochs
\citep{Brammer_etal_2009,Whitaker_etal_2011,Muzzin_etal_2013,Weaver_etal_2023}.
The measured quiescent fractions remain very large among massive galaxies in
clusters up to $z\sim 1$ \citep[e.g.][]{Mei_etal_2009,vanderBurg_etal_2020},
and quiescent galaxies are still found in the cores of massive systems up to
$z\sim 2$ albeit with a relatively large cluster-to-cluster variation
\citep{Andreon_etal_2014,Strazzullo_etal_2019,Willis_etal_2020}.

The recent advent of the James Webb Space Telescope (JWST) has significantly
advanced our ability to identify and characterize quiescent galaxies at very
early cosmic epochs. Thanks to its unprecedented sensitivity and resolution in
the near- and mid-infrared, deep surveys such as the Cosmic Evolution Early
Release Science programme (CEERS - \citealt{Finkelstein_etal_2023}), and the
JWST Advanced Deep Extragalactic Survey (JADES -
\citealt{Eisenstein_etal_2023}) have revealed a population of massive quiescent
galaxies that significantly outnumber expectations from theoretical models
\citep[][and references therein]{Carnall_etal_2023, Valentino_etal_2023,
  Baker_etal_2025a, Baker_etal_2025b}. While the number of spectroscopically
confirmed massive quiescent galaxies at early epochs increases steadily, their
environments remains poorly constrained from the observational point of view. A
few studies report evidence of local galaxy overdensities around massive
quiescent galaxies at $z\gtrsim 3$
\citep{Kubo_etal_2021,McConachie_etal_2022,Ito_etal_2023,Jin_etal_2024},
suggesting that at least a fraction of these can be found in overdense regions
that potentially trace the earliest cluster progenitors. Given the challenges
involved in a robust determination of the environment at these redshifts, it
remains unclear if these few existing case studies are representative of the
global population.

From a theoretical perspective, a strong correlation with environment would
align with expectations from a biased hierarchical formation scenario in which
early quenching occurs preferentially in the densest regions of the cosmic
web. However, it remains an open question whether environmental processes are
causally linked to quenching at these early times, or whether environment
merely correlates with early halo collapse while internal mechanisms dominate
the suppression of star formation. The latter scenario is more likely to apply
to the most massive galaxies, that are expected to experience only internal
physical processes for most of their lifetime \citep{DeLucia_etal_2012}.

A few recent studies have examined the environmental properties of quenched
high-z massive galaxies within theoretical models of galaxy formation, and the
results are somewhat contradictory.  \citet{Kimmig_etal_2025} use the
Magneticum Pathfinder\footnote{www.magneticum.org} simulation and find that the
36 quiescent galaxies with stellar mass larger than $3\times 10^{10}\,{\rm
  M}_{\odot}$ identified at $z\sim 3.4$ reside in underdense regions relative
to galaxies of similar stellar mass. \citet{Kimmig_etal_2025} argue that, in
the framework of the physical model for stellar and AGN feedback implemented in
Magneticum, this is a consequence of more efficient gas ejection and lack of
new gas replenishment in underdense regions. In contrast, \citet[][see also
  \citealt{Weller_etal_2025}]{Kurinchi-Vendhan_etal_2024}, using the
IllustrisTNG\footnote{https://www.tng-project.org/} simulations, find that the
$\sim 70$ massive\footnote{They select only galaxies with stellar mass larger
  than $10^{10.6}\,{\rm M}_{\odot}$} quiescent galaxies identified at $z\sim
3.7$ in the TNG100 and TNG300 volumes are preferentially located in more
massive halos and denser regions. These regions provide larger initial
reservoirs of infalling gas, which promotes black hole mass growth and leads to
quenching via AGN feedback\footnote{It is the radiatively inefficient regime
  that leads, in the framework of this model, to an efficient momentum
  injection by AGN feedback. TNG assumes a switch from the radiatively
  efficient to the inefficient mode that depends on the BH mass. Therefore, the
  faster black hole growth, promoted in over-dense regions, leads to earlier
  gas accretion in the inefficient regime.}. Similar trends are reported in
\citet{Szpila_etal_2025} for the SIMBA-C simulation.  These studies suggest
that investigating the environmental properties of massive quiescent galaxies
at early cosmic epochs can offer valuable insights into their formation
pathways, and into the physical processes shaping galaxy evolution in the early
Universe.

In this study, we carry out a throughout study of the local and large-scale
environment of massive quiescent galaxies at $3 < z < 5$ taking advantage of
our GAlaxy Evolution and Assembly (GAEA) model. This is a state-of-the-art
theoretical model that includes an updated treatment for both stellar and AGN
feedback and, as we have shown in previous work, reproduces the observed
quiescent fractions as well as number densities of massive quiescent galaxies
out to $z\sim 3-4$. It therefore represents an ideal tool for the detailed
study presented below. The layout of the paper is as follows: in
Section~\ref{sec:simsam}, we introduce the GAEA model and the simulation used
in this study. In Sections~\ref{sec:highzenv1} and \ref{sec:highzenv2}, we
characterize the environment of high-z quiescent model galaxies and in
Section~\ref{sec:highzprop} we analyze to what extent the physical properties
of high-z quiescent galaxies depend on their environment. In
Section~\ref{sec:future}, we characterize the future evolution of galaxies in
our model sample and its dependence on the environment. Finally, we summarize
our results and give our conclusions in Section~\ref{sec:discconcl}.

\section{The simulation and the galaxy formation model}
\label{sec:simsam}

GAEA\footnote{Details about the model, and access to a selection of data
  products, can be found at: https://sites.google.com/inaf.it/gaea/} is a
state-of-the-art theoretical model of galaxy formation that couples
prescriptions for the evolution of different baryonic components with
substructure based merger trees extracted from high-resolution cosmological
N-body simulations. The model builds on that presented in the original work by
\citet{DeLucia_and_Blaizot_2007}, but it has been updated significantly over
the past few years. Topical developments have been:
\begin{enumerate}
\item[(i)] a detailed treatment of the non-instantaneous recycling of gas,
  energy, and metals, which enables the tracing of individual metal abundances
  \citep{DeLucia_etal_2014};
\item[(ii)] an updated parametrization of stellar feedback partially based on
  results from hydro-dynamical simulations, tuned to reproduce the observed
  galaxy stellar mass function up to $z \sim 3$ \citep{Hirschmann_etal_2016};
\item[(iii)] a treatment for partitioning the cold gaseous phase of model
  galaxies in its atomic and molecular components, tuned to reproduce the
  measured HI and H$_2$ galaxy mass functions in the local Universe
  \citep{Xie_etal_2017};
\item[(iv)] a careful tracing of the angular momentum exchanges between
  different components, and a treatment for the non instantaneous stripping of
  the cold and hot gaseous components associated with satellite galaxies
  \citep{Xie_etal_2020};
\item[(v)] an improved model for cold gas accretion on super massive black
  holes (BHs) and AGN driven outflows that is tuned to reproduce the
  measured evolution of the AGN luminosity function up to $z\sim 4$
  \citep{Fontanot_etal_2020}.
\end{enumerate}

In previous work, we have shown that our model is able to reproduce a broad
range of observational measurements including the observed multiphase gas
content of central and satellite galaxies \citep{Xie_etal_2020}, the observed
secondary dependence of the local galaxy mass-gas metallicity relation
\citep{DeLucia_etal_2020}, and its evolution to higher redshift
\citep{Hirschmann_etal_2016,Fontanot_etal_2021}. Notably for this work, the
latest GAEA rendition \citep{DeLucia_etal_2024,Xie_etal_2024} reproduces nicely
the observed galaxy quenched fractions and the estimated number densities of
massive quiescent galaxies (when accounting for cosmic variance) up to $z\sim
3-4$. As discussed in \citet[][see also
  \citealt{DeLucia_etal_2025}]{DeLucia_etal_2024}, GAEA also performs
significantly better than other recently published theoretical models in
reproducing the estimated quenched fractions in galaxy clusters at $z\sim 1$.

The results presented in this paper are based on the Millennium Simulation
\citep{Springel_etal_2005}. This simulation follows 2,160$^3$ dark matter
particles in a box of 500~${\rm Mpc}\,{\rm h}^{-1}$ on a side, and assumes
cosmological parameters consistent with WMAP1 ($\Omega_\Lambda=0.75$,
$\Omega_m=0.25$, $\Omega_b=0.045$, $n=1$, $\sigma_8=0.9$, and $H_0=73 \, {\rm
  km\,s^{-1}\,Mpc^{-1}}$). More recent estimations provide slightly different
values for a few of these parameters, and in particular: a larger value for
$\Omega_m$ and a lower value for $\sigma_8$. However, we do not expect these
differences to affect significantly our model predictions and, in fact, our
recent work based on the P-Millennium Simulation \citep{Fontanot_etal_2025} has
shown that results from the latest rendition of our model converge very well
when run on different simulations from the Millennium suite \citep[see
  also][]{Wang_etal_2008,Guo_etal_2013}.

For the analysis presented below, we have selected massive quiescent galaxies
at $z \gtrsim 3.1$ and $z \lesssim 5.3$. Specifically, we have considered all
model galaxies in the redshift range of interest with stellar mass larger than
$10^{10.8}\,{\rm M}_{\sun}$\footnote{The stellar mass cut adopted is rather
  arbitrary and a compromise between the large estimated stellar mass quoted in
  recent observational studies \citep[e.g.][]{Carnall_etal_2024} and the goal
  to have a large enough statistical sample of model galaxies.}, after
accounting for an observational uncertainty of 0.25~dex, and with specific star
formation rate (sSFR) lower than $0.3\times t^{-1}_{\rm Hubble}$, where $t_{\rm
  Hubble}$ is the age of the Universe at the corresponding redshift. We note
that the assumed uncertainty is rather conservative at these redshift, and that
a different selection would not change qualitatively the results presented in
this work. Whenever appropriate, we will discuss in the following how results
depend on this convolution. We find a total of 4788 galaxies in the entire
Millennium Simulation box satisfying these criteria. We do not prevent the
descendant/progenitor of the same galaxies to be counted at different
snapshots. About 87 per cent of the selected galaxies are centrals and about 7
per cent are orphans, i.e. galaxies whose parent dark matter substructure has
been tidally stripped below the resolution limit of the simulation. In our
model, these galaxies survive for a residual merger time before merging (this
is justified by the fact that baryons are much more concentrated than dark
matter). Since the number density of massive quiescent galaxies decreases
rapidly with increasing redshift, the sample is dominated by galaxies at the
lowest end of the redshift range considered: 2195 galaxies are found at $z\sim
3.1$, 1287 at $z\sim 3.3$ and 668 at $z\sim 3.6$. There are only 12 massive
quiescent galaxies at $z\sim 5.3$ - the snapshot at the highest redshift
considered.

As a term of comparison for part of the analysis presented below, we also
consider a corresponding star forming sample, i.e. galaxies that are above the
same stellar mass cut but have sSFR larger than the threshold assumed for
quiescent galaxies. Our star forming sample is composed of 23,656 galaxies;
$\sim 88$ per cent of these are centrals and only $\sim 4$ per cent are
orphans.

\section{Are the first quiescent galaxies sitting in the most massive halos?}
\label{sec:highzenv1}

\begin{figure}
\centering
\resizebox{8cm}{!}{\includegraphics{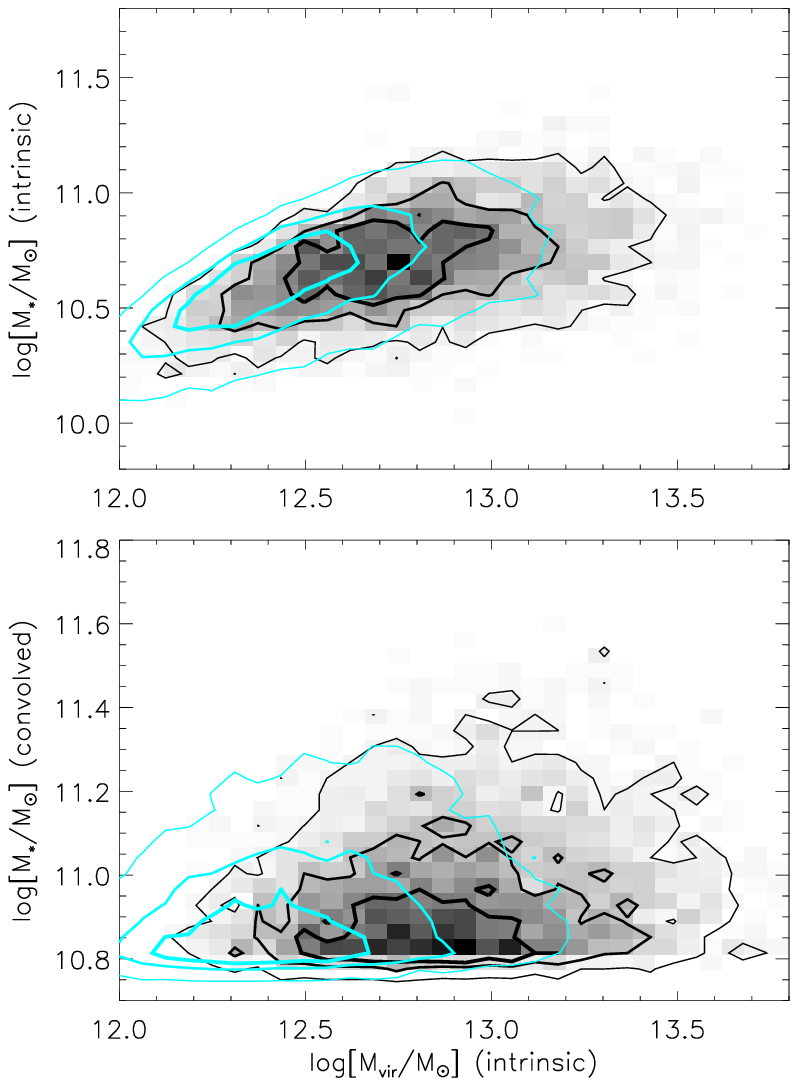}} 
\caption{Top panel: relation between stellar mass and parent halo virial mass
  for all galaxies selected at $z \gtrsim 3.1 $ and $z \lesssim 5.3$ (trends
  are similar when considering galaxies in smaller redshift ranges). Gray and
  cyan contours show the regions enclosing (from thicker to thinner lines) 30,
  60, and 90 per cent of the quiescent and star forming galaxies, respectively.
  Bottom panel: same as above but including an estimate of the observational
  uncertainty ($0.25$ dex) in the galaxy stellar mass.}
\label{fig:mstarmvir}
\end{figure}

The first question we address relates to which halos host the most massive
quiescent galaxies at $z \gtrsim 3$. The naive expectation (but see discussion
in Section~\ref{sec:intro}) is that these galaxies can be considered as the
signposts of significant overdensities in the early Universe, that might
eventually collapse in the most massive systems at lower redshifts. However,
the stellar-to-halo mass relation tends to be rather flat for halo masses
larger than $\sim 10^{13}\,{\rm M}_{\odot}$ \citep[][and references
  therein]{Girelli_etal_2020}, suggesting that a broader distribution of halo
masses should be expected. Fig.~\ref{fig:mstarmvir} shows the correlation
between galaxy stellar mass and parent halo mass for all model galaxies in our
samples. Gray and cyan contours show the regions enclosing (from thicker to
thinner lines) 30, 60, and 90 per cent of the quiescent and star forming
galaxies, respectively. To compute the distributions shown, we have combined
galaxies in the entire redshift range considered ($3 \lesssim z \lesssim 5.3$),
but we have verified that the trends are the same when considering narrower
redshift bins. The figure shows that massive quiescent galaxies reside in halos
that cover a broad range of masses, from Milky-Way like systems to the most
massive systems that can be identified at these high redshifts. In this sense,
they are not `special' when compared to star forming galaxies of similar mass,
that cover a similar range of parent halo masses. However, the distributions
obtained for star forming galaxies tend to be shifted to lower halo masses with
respect to those corresponding to quiescent galaxies, and the most massive
halos tend to host primarily quiescent galaxies. This is in agreement with
findings from \citet{Kurinchi-Vendhan_etal_2024} based on the IllustrisTNG
simulation, and can be understood as a consequence of an accelerated growth of
galaxies sitting at the center of very massive halos, and suppression of their
star formation activity either because of mergers or because of AGN
feedback. In our previous work, we have shown that the dominant quenching
channel for massive galaxies in GAEA is, in fact, AGN feedback
\citep{Xie_etal_2024}. Similar conclusions have been reached by independent
studies based on both hydro-dynamical simulations and semi-analytic models
\citep{Kurinchi-Vendhan_etal_2024,Szpila_etal_2025,Lagos_etal_2025}, with the
exception of the study by \citet{Kimmig_etal_2025} mentioned above and based on
the Magneticum simulation.

Fig.~\ref{fig:mstarmvir} shows that there is a weak correlation between the
intrinsic galaxy stellar mass and parent halo mass (top panel), both for
quiescent and star forming massive galaxies. However, such a correlation is completely
washed out when convolving the intrinsic galaxy mass with a (rather optimistic)
estimate of the observational uncertainties (bottom panel). Being rather flat
as a function of virial mass, the distributions shown in the bottom panel of
Fig.~\ref{fig:mstarmvir} do not vary significantly when further including an
estimate for the observational uncertainty on the parent halo mass.

\begin{figure}
\centering \resizebox{9cm}{!}{\includegraphics{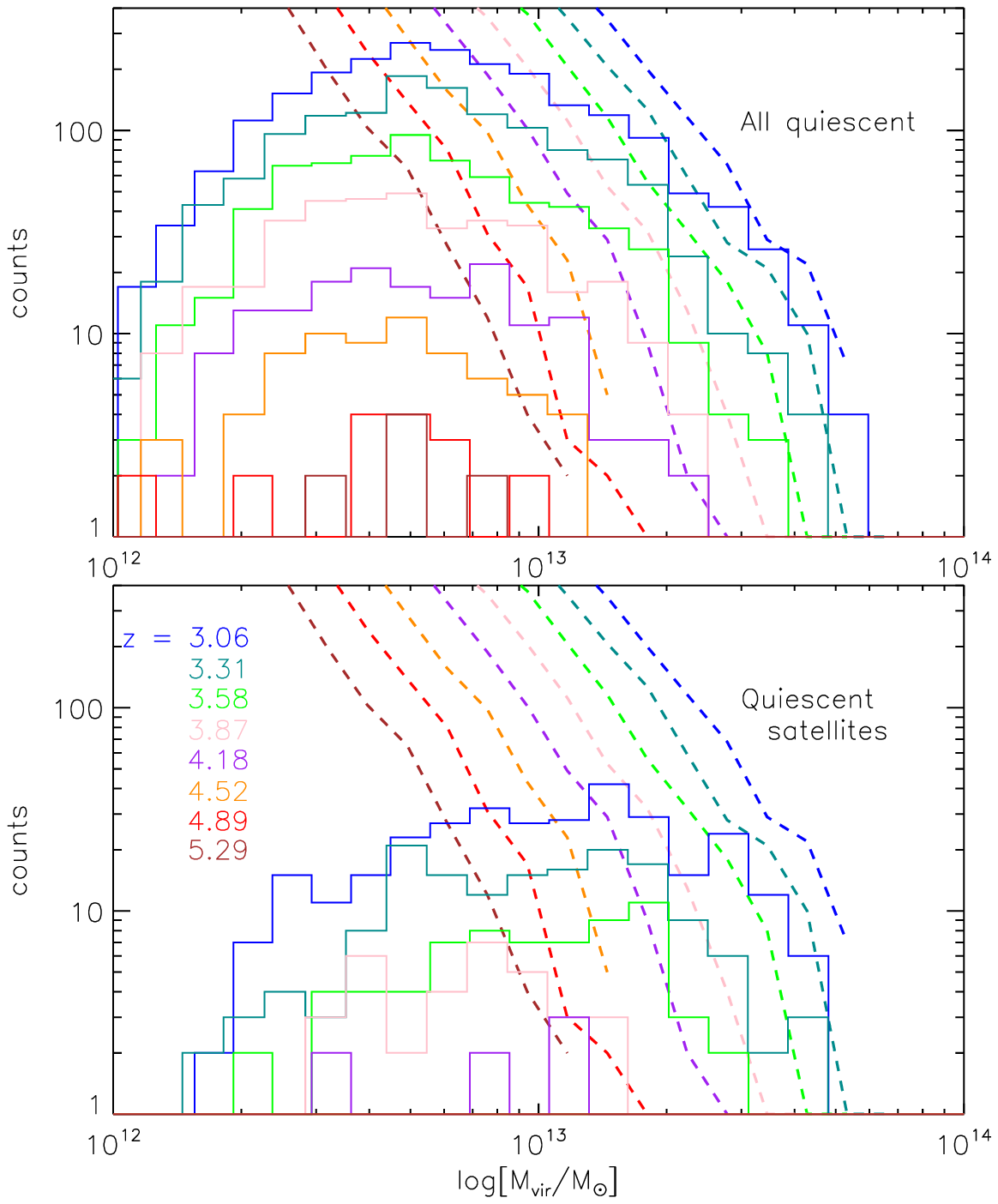}}
\caption{Halo mass distribution for quiescent model galaxies (solid lines) and
  the corresponding halo mass function (dashed lines). Different colors
  correspond to different redshifts (as indicated in the legend in the bottom
  panel), ranging from $\sim 3.1$ (blue) to $\sim 5.3$ (brown). The top and
  bottom panels are for all quiescent galaxies and for quiescent satellites
  only, respectively.}
\label{fig:halomf}
\end{figure}

\begin{figure*}
\centering
\resizebox{18cm}{!}{\includegraphics{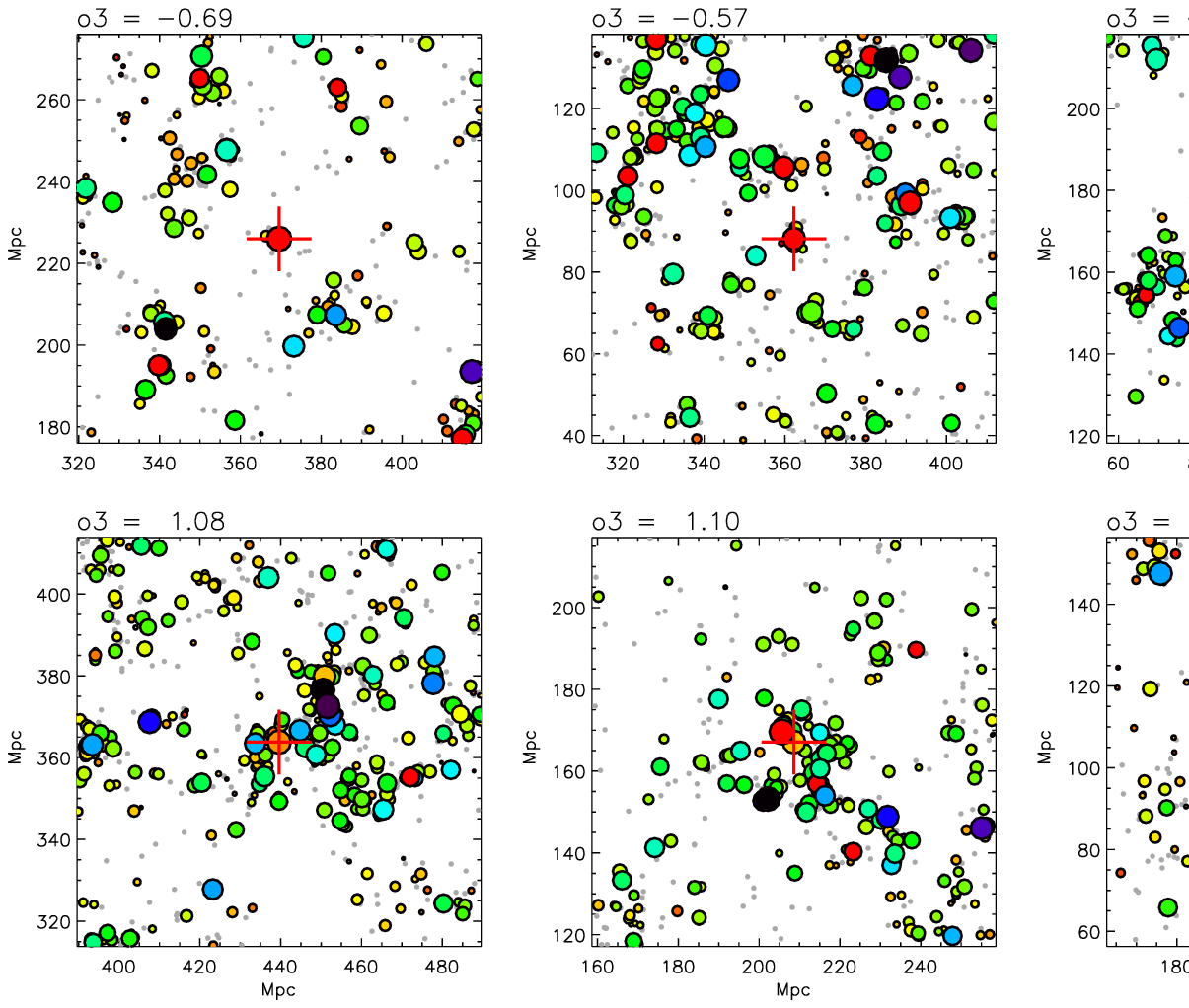}} 
\caption{Spatial distribution of galaxies selected at $z\sim 4.5$ and with
  over-densities over a 3~Mpc scale in the lowest 15th (top panels) and highest
  85th (bottom panels) percentiles of the distribution. Galaxies in our sample
  of massive quiescent galaxies are at the center of each panel and are marked
  by a red cross, while the other symbols mark the position of all galaxies
  more massive than $10^{9}\,{\rm M}_{\sun}$ in a projected region of 100x100
  Mpc comoving with a dept of $20$~Mpc comoving. The o$3$ value corresponding
  to each field is indicated in the top left corner of each box.  The color of
  the symbols scales with the galaxy star formation rate as indicated by the
  color bar on the right, while symbol size scales with galaxy stellar mass. We
  note that we have plotted the most massive galaxies at the end to emphasize
  their position, so they can hide other galaxies that lie close in
  projection.}
\label{fig:xypos}
\end{figure*}

Fig.~\ref{fig:halomf} shows, for all cosmic epochs considered, how the
distributions of halo masses corresponding to the massive quiescent galaxies in
our sample (solid lines) compare to the halo mass functions at the same cosmic
epoch (dashed lines). Different colors correspond to different redshifts, from
$\sim 3.1$ (blue) to $\sim 5.3$ (brown). The top and bottom panels are for all
quiescent galaxies and for quiescent satellite galaxies in our sample,
respectively. The figure confirms that, while the most massive halos at each
cosmic epoch tend to host central quiescent galaxies, many central galaxies in
halos more massive than $\sim 10^{13}\,{\rm M}_{\sun}$ are forming stars at non
negligible rates and therefore are not counted in the solid histograms in the
top panel of Fig.~\ref{fig:halomf}. The few satellite galaxies ($\sim 620$ out
of 4788) in our sample do not exclusively reside in the most massive halos but
are distributed in the entire halo mass range covered by our sample. Combining
the two panels, the figure also shows that the fraction of quiescent galaxies
increases with halo mass, as observed at least in the local Universe \citep[for
  a quantitative comparison between our model predictions and observational
  data, see][]{DeLucia_etal_2019}, and that this is not simply due to an
increase of the satellite fraction.

\section{What is the environment of high-z quiescent galaxies?}
\label{sec:highzenv2}

A measurement of parent halo mass is generally not easily accessible, in
particular for the massive quiescent systems that are being observed at
$z>3$. In this section, we introduce different estimates that might be more
easily obtained from observational data. In particular, for each galaxy in our
sample, we compute:

\begin{itemize}
\item[(i)] a `local density' estimate corresponding to the density of galaxies
  inside the projected circular region enclosing the 3 closest neighbors. We
  refer to this density as n$3$ in the following;
\item[(ii)] an `over-density' estimate corresponding to the mean over-density
  of galaxies measured within a projected circular region of radius equal to
  3~Mpc. This is normalized to the average value obtained at each redshift, by
  randomly drawing 1000 spheres with radius equal to 3~Mpc in the entire volume
  of the simulation. We refer to this over-density as o$3$ in the following.
\end{itemize}

In both cases, we have considered projected physical distances and only
galaxies in slices of depth $\sim 20~$Mpc with stellar mass larger than
$10^{9}\,{\rm M}_{\sun}$. We have verified that the results discussed below do
not vary significantly when increasing the number of neighbors (up to 5 or 7)
or the scale considered (we have considered overdensities on scales of 5 and
8~Mpc).

For illustrative purposes, we show in Fig.~\ref{fig:xypos} the projected
spatial distribution of eight galaxies, all at $z\sim 4.5$, from our sample of
quiescent galaxies. The galaxies have been selected to emphasize the variety of
environments in which model high-z quiescent galaxies live. The top panels
correspond to four galaxies in the lowest 15th percentile of the distribution
of the over-densities over 3~Mpc, while the bottom panels correspond to
galaxies in the highest 85th percentiles of the same distribution. Selected
galaxies are marked by a red cross and are shown at the center of the boxes,
that correspond to a depth of $20$~Mpc along the line of sight. We show the
projected distribution of all galaxies more massive than $10^{9}\,{\rm
  M}_{\sun}$ at the same redshift; symbol size scales with galaxy stellar mass
while color scales with the galaxy star formation rate, as indicated by the
color bars on the right of the figure. We note that we have plotted the most
massive galaxies at the end, so they might hide smaller galaxies at close
projected distances - this happens for example in the third bottom panel where
the massive quiescent galaxy at the center of the box is behind another massive
star forming galaxy (see Figs.~\ref{fig:xyposlow} and \ref{fig:xyposhigh} in
Appendix~\ref{sec.app1} for a similar plot, but showing only the distribution
of the galaxies with low and high star formation rates in the same
regions). The mean overdensity, which is indicated in the top left corner of
each box, is increasing from left to right in each row.

\begin{figure*}
\centering
\resizebox{18cm}{!}{\includegraphics{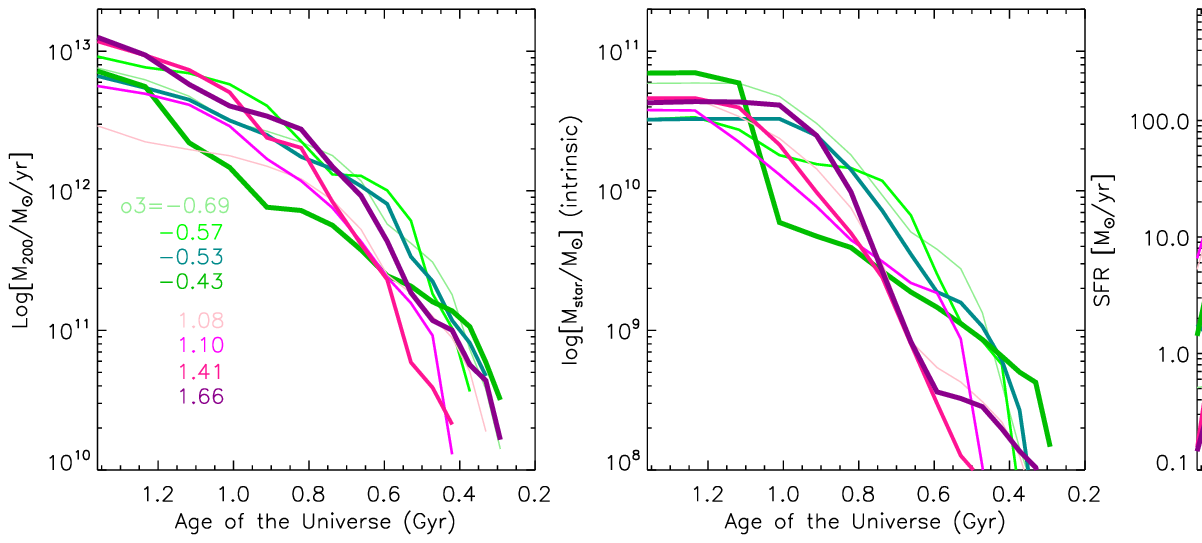}} 
\caption{Evolution of the parent halo mass, galaxy stellar mass, star formation
  rate, and black hole mass for the massive quiescent galaxies at $z\sim 4.5$
  considered in Fig.~\ref{fig:xypos}. Green and magenta lines refer to the top
  and bottom panel respectively, with thickness of the lines and color shade
  increasing with over-density values as indicated in the legend included in
  the left panel. Different line styles are used to show different galaxies for
  clearer visualization. $\rm {M}_{200}$ and ${\rm M_{BH}}$ are shown for the
  main progenitor at each previous cosmic epoch, while for the galaxy stellar
  mass and SFR we show the sum of all progenitors at the corresponding epoch.}
\label{fig:histexamples}
\end{figure*}

\begin{figure*}
\centering
\resizebox{18cm}{!}{\includegraphics{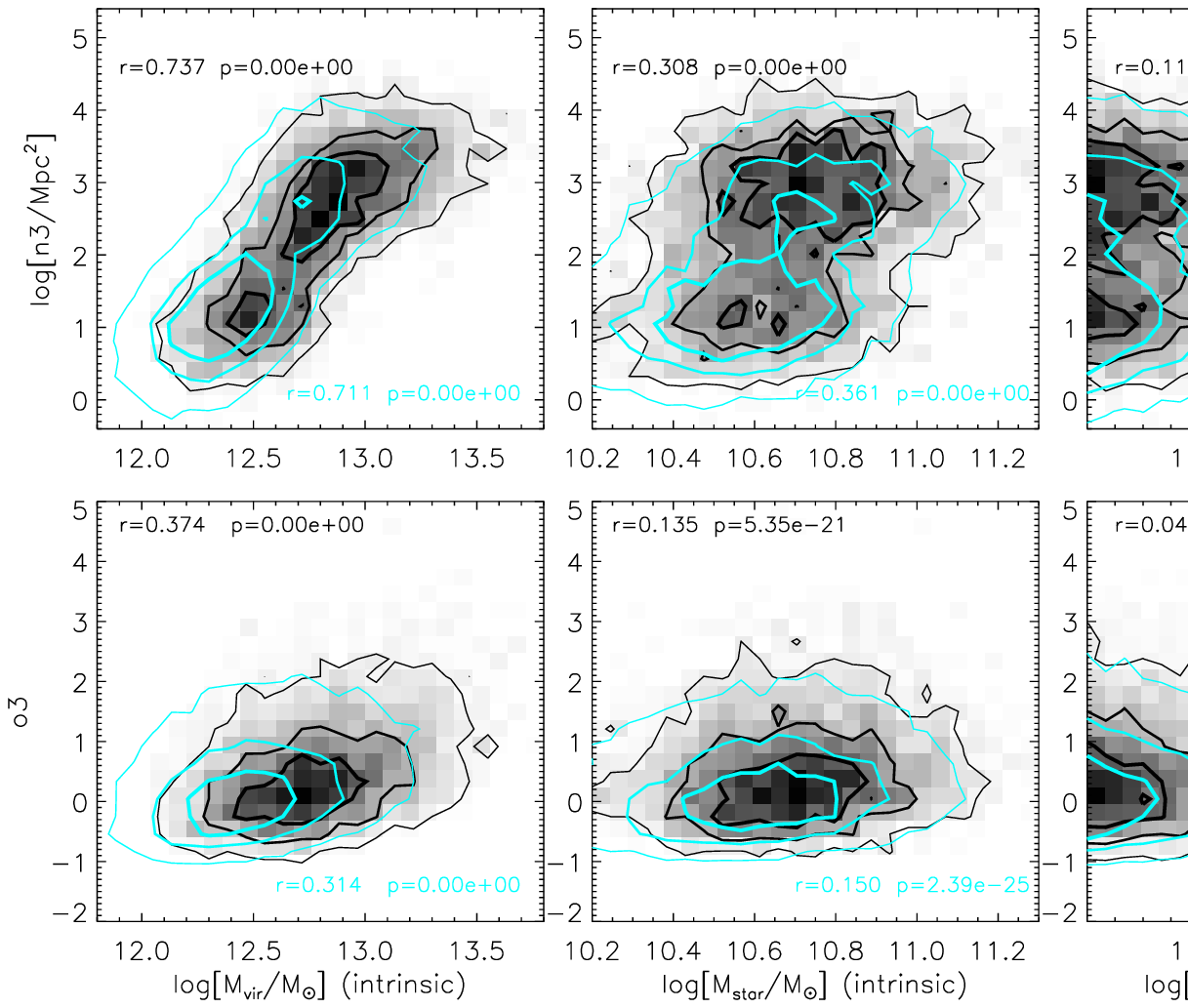}} 
\caption{Correlation between two different estimates of the galaxy environment
  and the parent halo mass (left column), the intrinsic galaxy stellar mass
  (middle column), and the galaxy stellar including an estimate for its
  uncertainty (right column). The top panels are for n$3$, i.e. the local
  density (in physical Mpc$^{-2}$) estimated considering the three closest
  neighbors for each galaxy in our quiescent (black) and star forming (cyan)
  samples. The bottom panels are for o3, i.e. the over-density estimated
  considering a scale of 3~Mpc. The numbers in each panel correspond to the
  Spearman's correlation coefficient and relative significance (a small value
  indicates a significant correlation) for quiescent galaxies (top left) and
  for star forming galaxies (bottom right).}
\label{fig:envscaling}
\end{figure*}

Fig.~\ref{fig:xypos} shows that there is a wide range of environments in which
early massive quiescent galaxies live, ranging from regions in which the
closest massive galaxies reside at projected distances larger than $\sim 20$
Mpc from the quiescent galaxy under consideration (top left panel) to regions
with many other massive galaxies at relatively small projected distances from
the target galaxy (e.g. bottom first and third panels).  In a few cases,
massive quiescent galaxies appear to reside on clear filamentary structures
(e.g. third top panel), towards the center of a void-like structure
(e.g. second top panel), or at the knot connecting different filaments
(e.g. the bottom right panel - incidentally, this region contains two of the
massive quiescent galaxies selected, the second being in the top right corner
of the projected box). The star formation rates of the galaxies residing in the
projected regions considered ranges from very low levels consistent with zero
to $\sim 250\,{\rm M}_{\odot}{\rm yr}^{-1}$, with a non negligible number of
massive actively forming stars in each region (see Fig.~\ref{fig:xyposhigh}).

Although a direct comparison is difficult, the range of environments visible in
Fig.~\ref{fig:xypos} is much broader from that depicted in Fig.~4 of
\citet{Remus_and_Kimmig_2025}, showing the projected surface brightness of the
gas at z=3.4 with superimposed the position of the six most massive quiescent
galaxies in the Magneticum Simulation Box3. In this simulation, massive
quiescent galaxies at high-z do inhabit different regions of the cosmic web as
is also in the case in our model, but none of them is found at the center (or
close to the center) of the hottest regions (the most massive halos). As
discussed in Section~\ref{sec:intro}, this qualitative difference in the
spatial distribution of massive quiescent galaxies is interesting because it
suggests that it can be used, in principle, to constraint different physical
mechanisms driving galaxy quenching at high-z.

As discussed in previous work \citep[][see also
  \citealt{Lagos_etal_2025}]{Xie_etal_2024}, in our model, the star formation
in massive galaxies is suppressed by AGN driven winds that are primarily driven
by mergers. Fig.~\ref{fig:histexamples} shows the evolution of the parent halo
mass, galaxy stellar mass, star formation rate, and black hole mass for the
same massive galaxies at $z\sim 4.5$ considered in Fig.~\ref{fig:xypos} (green
and magenta lines are used for the lowest and highest percentiles of the o$3$
distribution, with line thickness and color shade increasing with increasing
overdensity value). The figure shows that, already in this relatively small
sample of galaxies, there is a large variety of evolutionary histories. The
most massive galaxies do not necessarily reside in the most massive halos at
the epoch of observation, as already discussed in the previous section,
although the stellar mass assembly history parallels that of the parent dark
matter halos as expected. The most striking trend as a function of over-density
is that galaxies residing in the least over-dense regions tend to sit in halos
that have started growing earlier than their counter-parts residing in regions
that are over-dense at the epoch of the observation. The earlier stellar/halo
mass growth translates into an earlier growth of the black hole mass that can
lead to an earlier quenching of the star formation. Therefore, in this
sub-sample of galaxies, lower over-densities are explained by an accelerated
growth of halo/stellar mass before the epoch of the observations. It is
important to stress that these galaxies have not been selected randomly, but to
emphasize the variety of the environments at these early cosmic epochs. We will
show later that the trend just discussed becomes insignificant when considering
large samples of galaxies and large-scale overdensities as a proxy for the
environment. However, it remains significant when considering local densities.

A more quantitative characterization of the environment of the quiescent
galaxies in our sample is given in Fig.~\ref{fig:envscaling} that shows the
correlation between two different estimates of the galaxy environment and the
parent halo mass (left column), the intrinsic galaxy stellar mass (middle
column), and the same quantity convolved with an uncertainty of $0.25$~dex
(right column). The top panels correspond to the local density estimated
considering the three closest neighbors for each model galaxy (n$3$), while the
bottom panels correspond to an over-density estimated over a $3$~Mpc scale
(o$3$). Black and cyan lines show the distributions for quiescent and star
forming galaxies, respectively. The numbers in the top-left and bottom-right of
each panel correspond to the Spearman's correlation coefficient and relative
significance (a small value indicates a significant correlation) for the two
galaxy samples. The top-left panel shows that there is a good although broad
correlation between n$3$ and the parent halo mass, both for star forming and
quiescent galaxies. A weak correlation is also visible between n$3$ and the
intrinsic stellar mass (top-middle panel): the most massive galaxies tend to
have larger local densities, but the scatter is very large and large values of
n$3$ can be found for a quite broad range of galaxy stellar masses. This
already weak correlation is almost entirely washed out when including an
(optimistic) estimate of the uncertainty on the galaxy stellar mass (top-right
panel).  The correlations with o$3$ are overall weaker and an uncertainty on
galaxy stellar mass can even invert the correlation with galaxy stellar mass
(bottom-right panel). The trends just discussed are very similar for quiescent
and star forming galaxies, but for a small shift towards lower halo and stellar
masses for the star forming galaxies. This shift appears to be more evident
when considering local densities than larger scale over-densities. Results are
qualitatively similar when considering local densities based on a larger number
of neighbors or over-densities computed on larger scales.
\section{Do physical properties of high-z quiescent galaxies depend on the environment?} 
\label{sec:highzprop}

\begin{figure}
\centering
\resizebox{8cm}{!}{\includegraphics{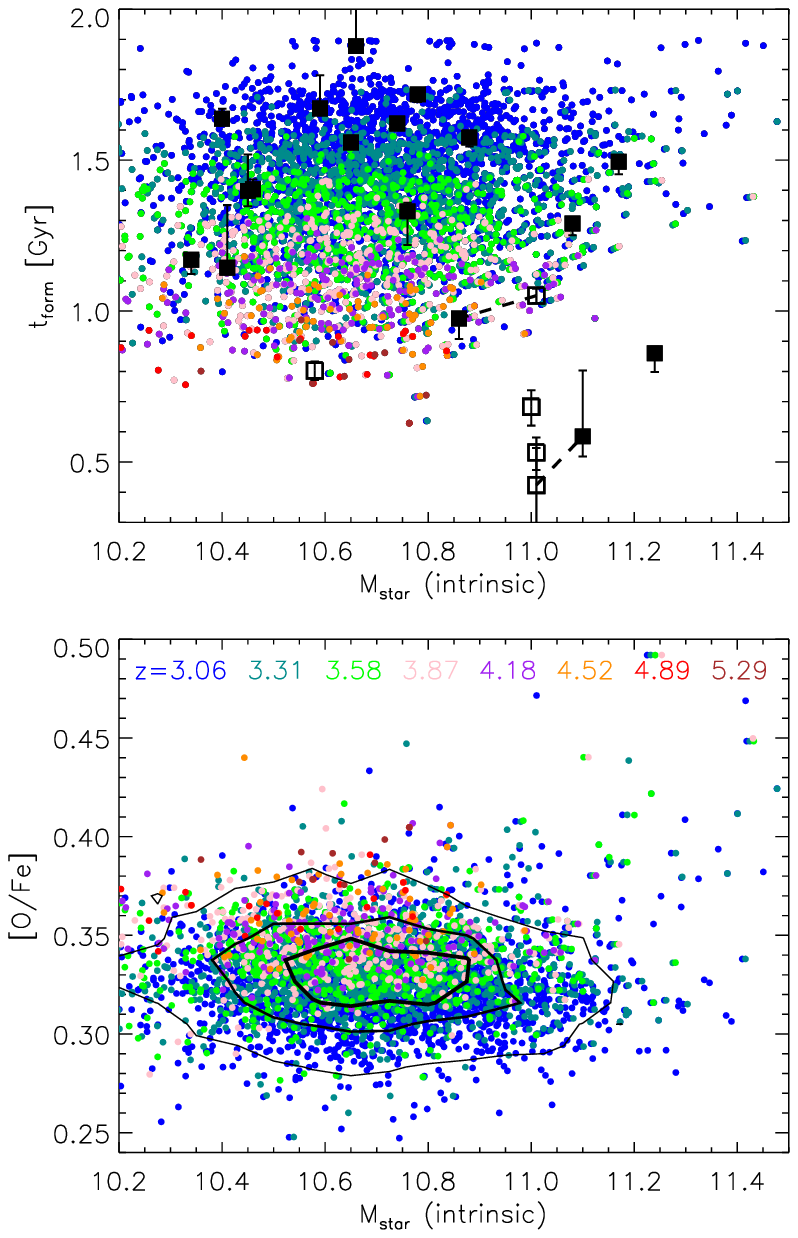}} 
\caption{Formation time (top panel) and [O/Fe] (bottom panel) as a function of
  galaxy stellar mass. The former is defined as the age of the Universe at
  which half of the stars in the galaxies are formed. Color coding is as a
  function of redshift, as indicated in the legend in the bottom panel. Open
  and filled symbols with error bars in the top panel correspond to
  observational measurements by \citet[][5 galaxies between $z\sim 3.2$ and
    $z\sim 4.6$]{Carnall_etal_2024} and \citet[][19 galaxies with a median
    redshift of $\sim 3.55$]{Nanayakkara_etal_2025}, respectively.  Contours in
  the bottom panel show the regions enclosing (from thicker to thinner) 30, 60,
  and 90 per cent of the model sample.}
\label{fig:props}
\end{figure}

Having characterized the environments in which massive quiescent galaxies live,
we can then ask to what extent their physical properties depend on their local
or global environment. In this section we focus on two different quantities for
which observational measurements are rapidly being collected and/or will
rapidly increase in the future: star formation histories and the stellar
metallicities.

\begin{figure*}
\centering
\resizebox{18cm}{!}{\includegraphics{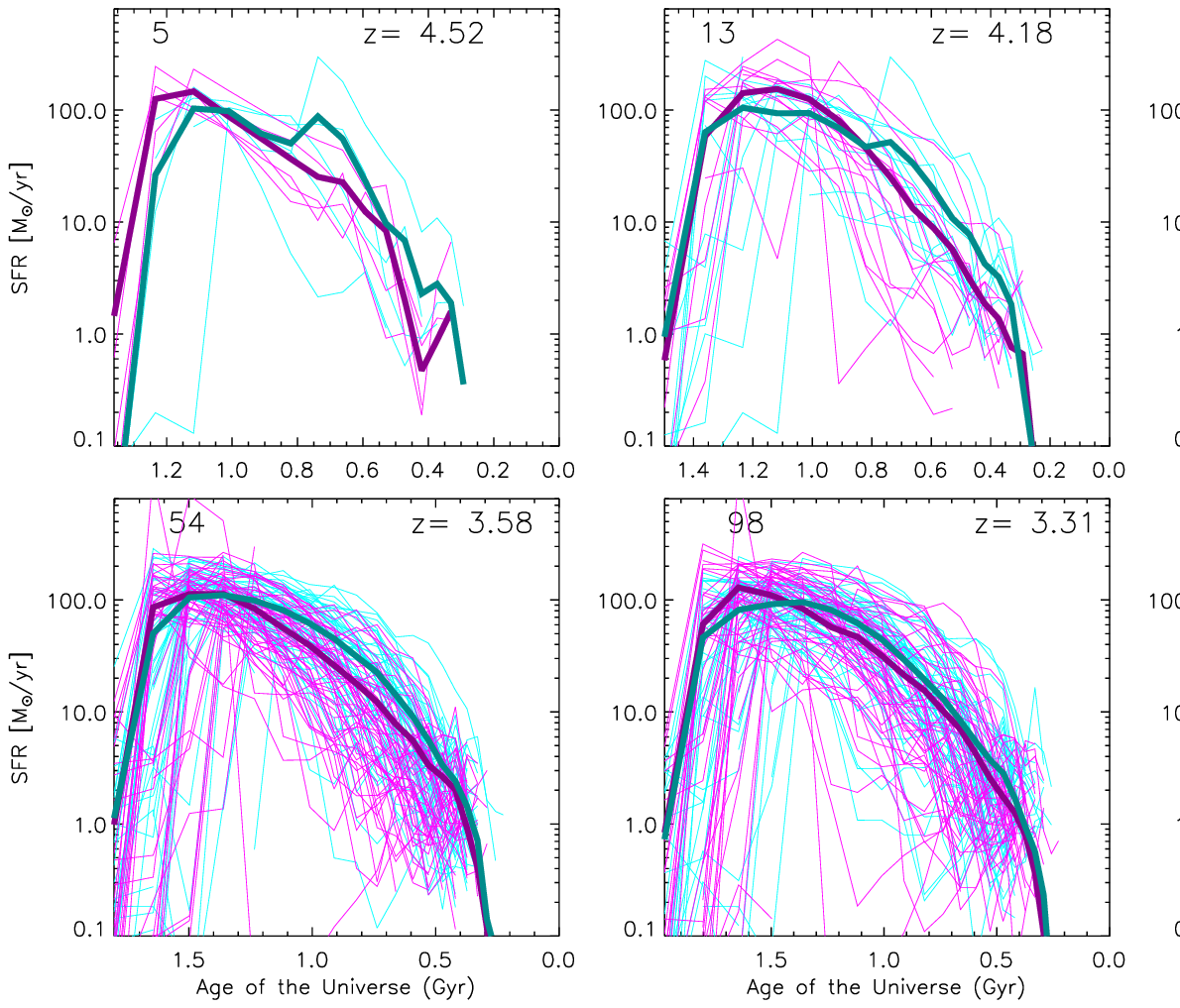}} 
\caption{The star formation histories of quiescent galaxies in the bottom 25th
  (cyan) and highest 75th (magenta) percentiles of the distribution of n$3$,
  considering matched distributions of parent halo mass. Thicker lines show the
  median star formation histories for each sub-sample. The numbers of galaxies
  in each sub-sample at each snapshot is given in the top left corner.  A
  maximum of 50 individual star formation histories in each sub-sample is shown
  to avoid overcrowding the panels.}
\label{fig:sfhextr}
\end{figure*}

Galaxies in our sample of model quiescent systems are old, metal-rich, and
$\alpha$-enhanced. Fig.~\ref{fig:props} shows the distributions of the
formation times, defined as the age of the Universe when half of the stellar
mass has been formed\footnote{For our model galaxies, this is computed by
  linearly interpolating the predicted stellar mass at the outputs available.},
and the [O/Fe] abundances as a function of galaxy stellar mass. The
color-coding indicates the redshift (as in Fig.~\ref{fig:halomf}). Open
symbols with error bars in the top panel correspond to the observational
estimates by \citet{Carnall_etal_2024} for 5 massive quiescent galaxies between
$z\sim 3.2$ and $z\sim 4.6$. Filled symbols correspond to measurements by
\citet{Nanayakkara_etal_2025}, based on a sample of 19 quiescent galaxies with
a median redshift of $\sim 3.55$. For all these galaxies, JWST/NIRSpec
spectroscopy is available. The two pairs of symbols connected by dashed lines
correspond to the galaxies: ZF-UDS-6496 at $z\sim 4$, with an estimated
formation redshift of $\sim 5.6$ in \citet{Carnall_etal_2024} and $\sim 5.9$ in
\citet{Nanayakkara_etal_2025}, and ZF-UDS-7329 at $z\sim 3.2$ with an estimated
formation redshift of $\sim 11.2$ in \citet{Carnall_etal_2024} $\sim 8.8$ in
\citet{Nanayakkara_etal_2025}. The range of estimated formation times for the
same systems clearly shows the large systematic uncertainties, potentially more
relevant than the reported statistical errors, associated with these
measurements. One obvious source of systematic uncertain is a different
(e.g. more top-heavy) stellar initial mass function - this is something that we
plan to address in future work.

Our model predictions cover quite well the entire range of formation times
inferred from the data. Note that, in Fig.~\ref{fig:props}, we show on the
x-axis the predicted intrinsic stellar masses to account for larger accuracy of
the estimates based on spectroscopic information. However, even with
spectroscopy observational uncertainties are not negligible and, accounting for
them, many points would be moved to the right of the figure. At least a few
would end up close to the massive galaxies in the observed sample with very
high formation redshifts. The lower panel of Fig.~\ref{fig:props} shows that
all model quiescent galaxies in our sample are $\alpha$-enhanced, with no
significant trend as a function of stellar mass but a trend as a function of
redshift: quiescent galaxies identified at higher redshifts are more
$\alpha$-enhanced, as expected given the shorter formation times. To date, the
only available measurement of $\alpha$-enhancement at these high redshifts has
been reported by \citet{Carnall_etal_2024}: for the galaxy ZF-UDS-7329 at
$z\sim 3.2$, they report a value of [Mg/Fe]$=0.42^{+0.19}_{-0.17}$. Considering
an average positive offset of $\sim 0.10-0.15$ between [O/Fe] and [Mg/Fe], this
estimate is significantly higher than what is predicted for the bulk of our
model galaxies at the same redshift, indicating that for most of the model
galaxies star formation occurs over a longer time-scale than inferred for this
particular system. The relatively large uncertainties on these observational
measurements ease concerns about the disagreement, and a top-heavy IMF would
also lead to larger [$\alpha$/Fe] ratios for model galaxies. In addition, we
note that a few galaxies do exist in the simulated volume with intrinsic
stellar mass larger than $\sim 10^{10.8}\, {\rm M}_{\odot}$ and [O/Fe] larger
than $\sim 0.40$. Therefore, it is not impossible to have galaxies with
properties similar to those inferred for ZF-UDS-7329 in the framework of our
model.

Fig.~\ref{fig:sfhextr} shows the average star formation histories of the
galaxies in our sample in the 25th (dark cyan) and 75th (dark magenta)
percentiles of the distribution of n$3$. We have omitted the very few galaxies
in the highest redshift snapshots considered because of the low number
statistics, and we have matched\footnote{Due to the significant correlation
  between n$3$ and halo mass (top left panel of Fig.~\ref{fig:envscaling},
  there is still a systematic small difference between the median halo mass in
  the two distribution of $\sim 0.1$~dex.}  the distributions of parent halo
mass for the galaxies in the two extreme environments considered. The number of
galaxies used at each redshift and for each environment bin is shown in the top
left corner. The figure shows that there is a clear difference between galaxies
residing in the regions corresponding to the highest and lowest values of n$3$.
The trends reflect those described for Fig.~\ref{fig:histexamples}: star
formation tends to start earlier, proceed at larger rates, and declines earlier
in regions corresponding to the lowest percentiles of the n$3$ distribution. No
strong difference is observed for the times-scale of quenching, that is in all
cases very short (see Fig.~11 in \citealt{Lagos_etal_2025}). We have verified
that the differences evident in Fig.~\ref{fig:sfhextr} are largely unaffected
when considering local estimates based on a larger number of neighbors (n$5$
and n$7$). The different star formation histories corresponding to galaxies in
regions with different (local) densities do not translate into a systematic
difference in the ages of the stellar populations (when using the formation
time as a proxy) of galaxies residing in these different environments. This is
due to the very large scatter of the individual star formation histories as
indicated by the thin lines (to avoid overcrowding the panels, we show a
maximum of 50 - randomly selected - galaxies in each subsample at each
redshift). We have also verified that, when considering a global estimate of
the environment, the differences between the median star formation histories
corresponding to the two extremes is significantly reduced (see
Fig.~\ref{fig:sfhextro3} in Appendix~\ref{sec.app1}).

\begin{figure}
\centering
\resizebox{7cm}{!}{\includegraphics{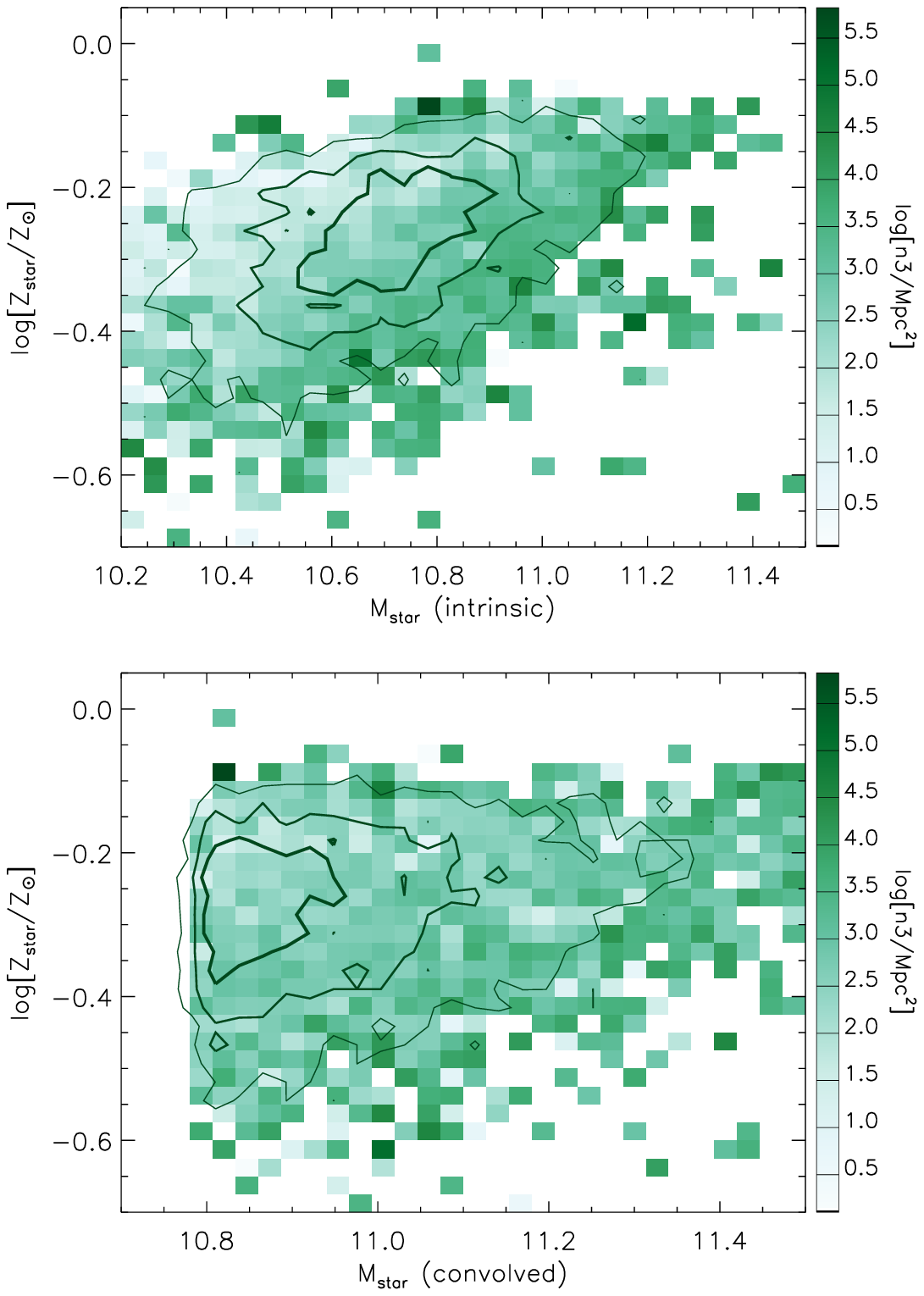}} 
\caption{Mass-(stellar)metallicity relation color-coded as a function of local
  density (n$3$). The upper panel shows the correlation between stellar
  metallicity and intrinsic galaxy stellar mass, while in the bottom panel
  model masses have been perturbed assuming an uncertainty of 0.25 dex on the
  galaxy mass.}
\label{fig:metrels}
\end{figure}

Fig.~\ref{fig:metrels} shows the mass-metallicity relation for all quiescent
galaxies in our sample, color coded as a function of log(n$3$). In the top
panel, the intrinsic galaxy stellar mass has been used, while in the bottom
panel the galaxy stellar mass includes an uncertainty of $0.25$~dex. The figure
shows that there is a very weak trend with local density that is more evident
for the galaxies with lowest mass in our sample: stellar metallicity is larger
for galaxies in regions of lower local density as a consequence of the more
prolonged star formation histories. However, this trend is completely washed
out when an (optimistic) uncertainty on the galaxy stellar mass is included. We
have verified that no significant trend is visible in case a global density
environment is considered. In absolute values, the available observational
estimates of stellar metallicity reach larger values than those shown in
Fig.~\ref{fig:metrels}, but cover a wide range and are affected by large
uncertainties: \citet{Nanayakkara_etal_2025} report values of stellar
metallicity ranging between $\sim -0.55$ and $\sim 0.54$; for three of the
galaxies in \citet{Carnall_etal_2024}, the estimated metallicity is $0.32-0.35$
while for the other two the estimated values are $-0.41^{0.06}_{-0.09}$ and
$-0.96^{0.04}_{-0.09}$. In fact, these measurements for high-redshift quiescent
galaxies are very challenging because of the very weak metal absorption
features due to young (in absolute terms) ages and of the highly
$\alpha$-enhanced stellar populations. We refrain to add observational
estimates to Fig.~\ref{fig:metrels} because of the very large estimated
uncertainties, but note that model predictions also cover a wide range of
metallicities.

\section{What are the descendants of early quiescent galaxies?}
\label{sec:future}

\begin{figure}
\centering \resizebox{8cm}{!}{\includegraphics{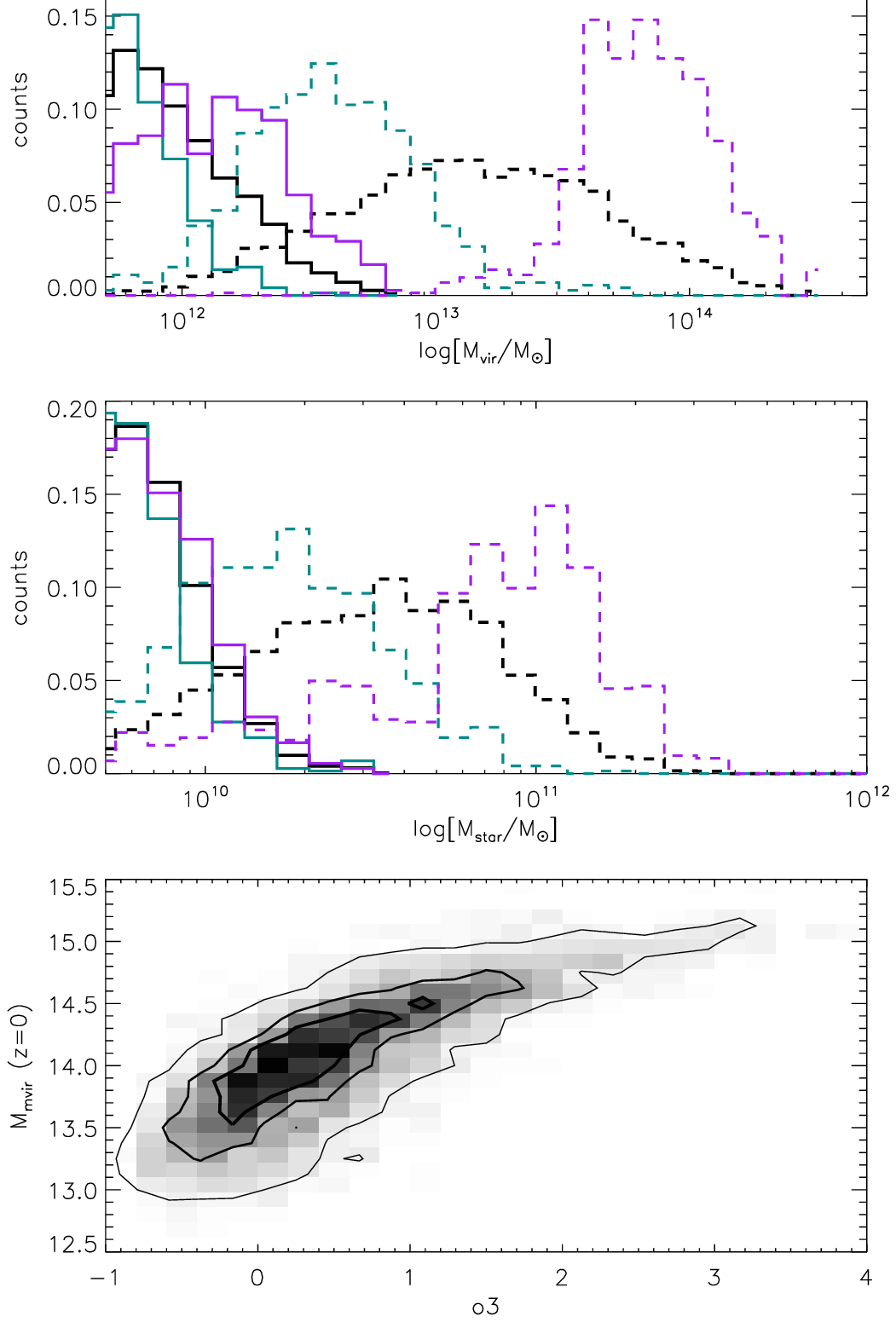}}
\caption{Solid histograms show the distribution of parent halo mass (top panel)
  and stellar mass (middle panel) for all quiescent model galaxies. Dashed
  histograms are the corresponding distributions for their descendants at
  $z=0$. Cyan and magenta are used for galaxies that are in the extremes (15th
  and 85th percentiles) of the distribution of o3. The bottom panel shows the
  correlation between the virial mass of the halos in which descendants of
  galaxies in our sample reside at $z=0$ and the over-density at the time of
  the observation. Contours in the bottom panel show the regions enclosing
  (from thicker to thinner) 30, 60, and 90 per cent of the galaxies.}
\label{fig:descmass}
\end{figure}

One last question we can ask, taking advantage of our model results, is related
to what are the descendants of high-z quiescent galaxies and if their future
fate depends on the environments in which they reside. As noted above, about 88
per cent of the galaxies selected at high-z are centrals. Tracing their
descendants down to $z=0$, we find that about 77 per cent are still centrals
while about 11 per cent are orphans. Fig.~\ref{fig:descmass} shows the
distribution of parent halo mass (top panel) and stellar mass (middle panel)
both for the galaxies selected at high redshifts (solid lines) and their
descendants at $z=0$ (dashed histograms). The purple and cyan lines correspond
to the galaxies that are in the extremes (15th and 85th) percentiles of the
distribution of o$3$ at high-z. The bottom panel shows the correlation between
the parent halo mass at $z=0$ and the over-density o$3$ at the time of
observation. The figure shows clearly that the distributions of halo masses and
stellar masses of the $z=0$ descendants of high-z quiescent galaxies are very
broad, broader than the distribution found at the redshift of observation. This
can be understood as a natural consequence of the stochasticity of merger
events. The effect, that is stronger the higher the redshift, has been noted in
earlier work and recognized as a generic prediction of hierarchical growth
\citep[e.g.][]{DeLucia_and_Blaizot_2007,Trenti_etal_2008,Angulo_etal_2012}.
Fig.~\ref{fig:descmass} also shows that the large-scale overdensity has a
significant impact on the fate of the high-z quiescent galaxies: galaxies that
are sitting in the largest overdensities have higher probabilities to end up in
a massive cluster today. There is a clear and rather strong correlation between
the over-density at the time of the observation and the halo mass at $z=0$,
which makes the latter a good predictor of the environment in which the
descendants of high-z passive galaxies live at present.  A clear but less
significant difference appears when considering a local estimate of the
environment. The latter is a more powerful indicator of the halo mass at the
redshift of observations, but a less powerful indicator of the $z=0$ halo mass
(see Fig.~\ref{fig:descmassn3} in Appendix~\ref{sec.app1}). Similar results
have been found in the recent work by \citet{Remus_and_Kimmig_2025}. Their
Fig.~15 shows a broad correlation between more dense environments at $z=3.4$
and larger $z=0$ halo masses. However, 75 per cent of their high-z quiescent
galaxies lie in underdense environment according to their definition. As noted
above, this is likely related to how AGN feedback is implemented in their
simulation, and it is worth noting that their particular implementation of
stellar and AGN feedback does not lead to a good agreement with the observed
evolution of the galaxy stellar mass function at lower redshifts
\citep{Hirschmann_etal_2014b,Steinborn_etal_2015}.

\begin{figure*}
\centering
\resizebox{18cm}{!}{\includegraphics{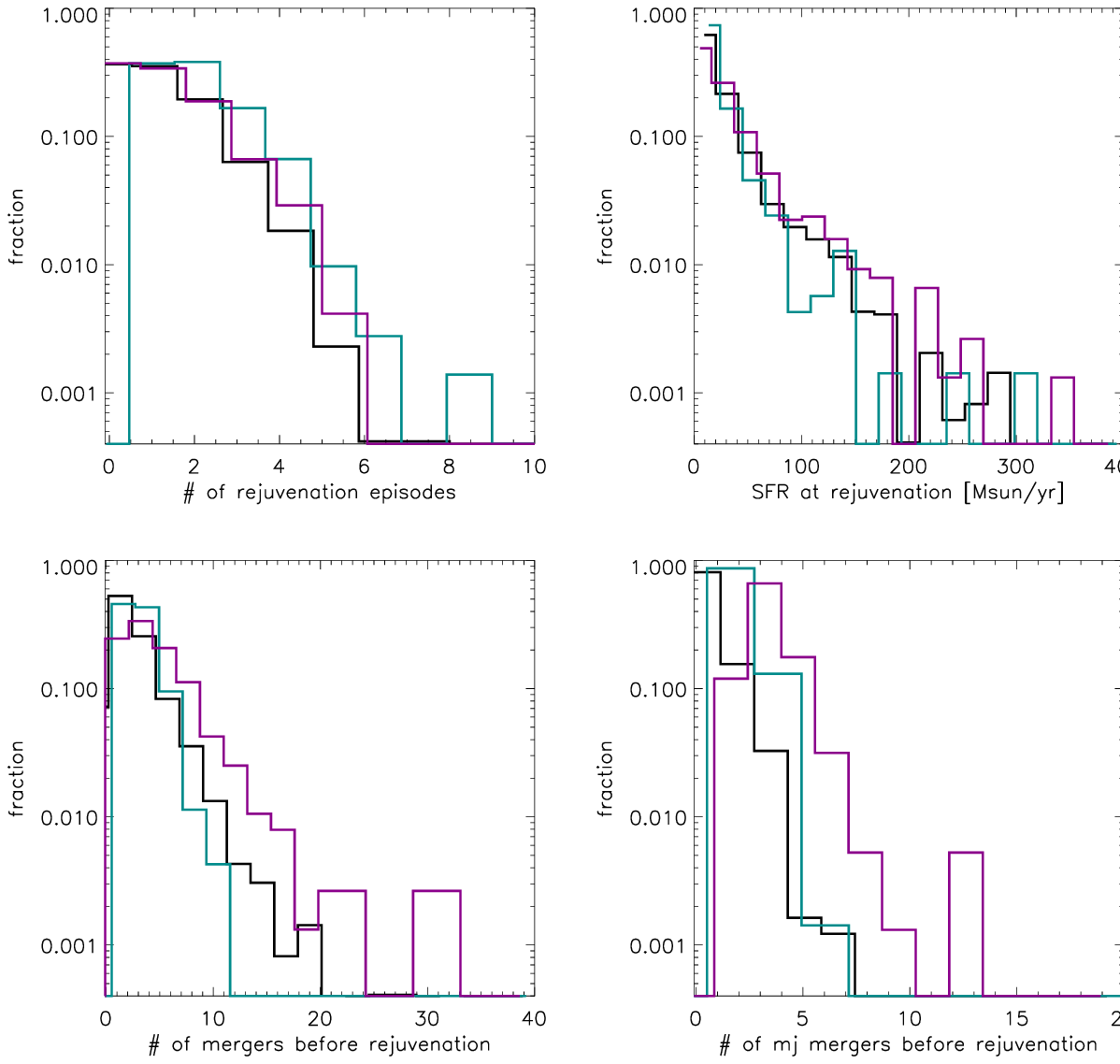}} 
\caption{Top panels: distributions of the number of rejuvenation episodes, of
  the SFR during each episode, and of the associated time-scale. Bottom panels:
  distributions of the total number of mergers in the last snapshot just before
  the rejuvenation episodes, of the number of mergers with a mass ratio of at
  least 1 to 10, and of the associated fractional increase of stellar mass
  (excluding the stellar mass that is accreted through mergers). Magenta and
  cyan are for the 15th and 85th percentiles of the o$3$ distributions.}
\label{fig:rej}
\end{figure*}

A more detailed characterization of the future star formation history of high-z
quiescent galaxies is shown in Fig.~\ref{fig:rej}. For each model galaxy, we
have followed the descendant to $z=0$ and identified all snapshots where the
sSFR is above the threshold used to identify a galaxy as quiescent ($0.3\times
t^{-1}_{\rm Hubble}$ - see Section~\ref{sec:simsam}). We have merged together
adjacent snapshots when counting these ``rejuvenation'' episodes. The top
panels of Fig.~\ref{fig:rej} show the distributions of: the number of
rejuvenation episodes, the SFR during each episode, and the associated
time-scale. The bottom panels show the distributions of: the total number of
mergers in the last snapshot just before the rejuvenation episodes, the number
of mergers with a mass ratio of at least 1 to 10, and the associated increase
of stellar mass (excluding the stellar mass that is accreted through
mergers). Magenta and cyan lines are for the 15th and 85th percentiles of the
o$3$ distributions, as in previous figures.

In the framework of our model, rejuvenation is not a common event in the
future lifetime of a massive quiescent galaxy at high-z. About 36 per cent of
the quiescent galaxies in the selected sample remain passive down to $z=0$,
experiencing no rejuvenation events. For the rest of the galaxies in the
sample, the mean number of rejuvenation episodes is $\sim 1.6$ (with a median
number of 1). Only very few galaxies experience a number of rejuvenation
episodes larger than 4 (15 galaxies in total out of 4788). The overall fraction
of galaxies experiencing rejuvenation in our model is larger than quoted in
\citet{Szpila_etal_2025} but very close to those estimated in
\citet{Remus_and_Kimmig_2025}. However, we note that these numbers are very
sensitive to the exact definitions adopted.  The galaxies with more frequent
episodes of rejuvenation events tend to reside in the less over-dense regions
at the time of observations, but the number statistics are very low.

About 34 per cent of the rejuvenation episodes are associated with a star
formation rate lower than $10\,{\rm M}_{\odot}\,{\rm yr}^{-1}$, but $\sim 5$
per cent of the galaxies in our sample have rejuvenation episodes with
associated rates larger than $100 \,{\rm M}_{\odot}\,{\rm yr}^{-1}$. A larger
number of these episodes are associated with galaxies that reside in the
over-dense regions at the epoch of observations. The vast majority of the
rejuvenation episodes correspond to events occurring over one snapshot only of
the simulation: the interval between subsequent snapshots varies between $\sim
110$ and $\sim 270$~Myr at $z\sim 5$ and $z=0$ respectively. Only a few
episodes last much longer, i.e. the galaxy accretes significant amount of gas
to sustain star formation over several snapshots. Rejuvenation episodes are
typically associated with recent mergers. A few of the galaxies in our sample
experience relatively large number of mergers, but many of these have very low
mass-ratio (see bottom left and bottom middle panel). For about 15 per cent of
the galaxies in the sample, the increase in mass associated with the
rejuvenation episodes (subtracting the mass accreted through mergers) is non
negligible (larger than $10^{10}\,{\rm M}_{\odot}$). The fractional mass
increase with respect to the present stellar mass tends to be larger for
galaxies in underdense environments (bottom-right panel), due to the large
difference between the present day stellar mass of galaxies residing in
overdense and underdense environments (see middle panel of
Fig.~\ref{fig:descmass}).

\section{Summary and conclusions}
\label{sec:discconcl}

In this work, we have analyzed the environment of massive quiescent galaxies at
$3 \lesssim z \lesssim 5$ using our GAlaxy Evolution and Assembly (GAEA)
theoretical model of galaxy formation. We selected galaxies with stellar mass
larger than $10^{10.8}\,{\rm M}_{\sun}$ and specific star formation rate below
$0.3\times t^{-1}_{\rm Hubble}$, resulting in a sample of about $5,000$
galaxies within a simulated volume of $\sim 685$~Mpc comoving on a side. In
previous work, we have shown that the latest rendition of our model reproduces
a wide range of observational results, including the observed quiescent
fractions and number densities up to $z\sim 3-4$, providing an ideal tool for
the analysis presented in this paper. We show that the formation times derived
for model galaxies at $z\gtrsim 3$ cover well the range inferred from recent
observational work \citep{Carnall_etal_2024,Nanayakkara_etal_2025}. Galaxies
with very short formation time-scales and early formation epochs are present in
our model, though they are rare. Systems like ZF-UDS-7329 are therefore not
impossible in the framework of our model, though their true abundance in the
Universe remains uncertain.

Model high-z quiescent galaxies are $\alpha$-enhanced and exhibit a wide range
of stellar metallicity broadly consistent with observational trends, though
they do not reach the highest metallicities recently inferred. These
measurements remain challenging due to the weakness of diagnostic spectral
lines and the lack of well tested $\alpha$-enhanced stellar population models
for young ($\lesssim 1$~Gyr) ages. More accurate observational constraints will
be essential to use these chemical signatures to test and refine current galaxy
formation model.

Our main conclusions can be summarized as follows:

(i) {\it Environmental diversity}: massive quiescent high-z galaxies inhabit a
wide range of environments, spanning from underdense regions resembling voids,
to filaments, walls, and dense knots at the intersections of filaments. Their
environment, whether characterized by halo mass, local neighbors counts, or
large-scale overdensity, are far from uniform.

(ii) {\it Environmental trends and halo history}: quiescent galaxies that live
in under-dense regions at the epochs of the observation tend to reside in halos
that collapsed earlier and experienced rapid early growth. This trend is most
evident when considering a local estimate for the environment, and becomes less
clear when using large-scale overdensity. In addition, these correlations can
be diluted or even inverted if halo mass is not controlled for, making their
observational detection challenging given current uncertainties and the
intrinsic large scatter in star formation histories.

(iii) {\it Descendant diversity}: the descendants of high-z massive quiescent
galaxies span a broad range of halo and stellar masses by $z=0$, due to the
stochastic nature of mergers. Some galaxies evolve little and remain in
group-size halos, while others grow significantly and end up in cluster-size
halos by $z=0$. The large-scale overdensity measured at the epoch of the
observations serves as a useful, though not perfect, predictor of the
descendant halo mass at $z=0$.
  
(iv) {\it Rejuvenation}: about 36 per cent of the massive quiescent galaxies in
our sample do not experience any subsequent rejuvenation, and only a few
galaxies undergo multiple ($>4$) rejuvenation events accompanied by strong
bursts of star formation. Most rejuvenation episodes are merger-driven and
occur more frequently in overdense environments at the epoch of observation.

These findings have important implications for current and future observational
programs aimed at using environmental diagnostics to test quenching mechanisms
or to identify the progenitors of present-day massive clusters. Both the number
densities and environmental properties of massive quiescent galaxies at
$z\gtrsim 3$ are expected to exhibit significant cosmic variance. Furthermore,
such galaxies do not necessarily reside in the most massive halos of their
epoch, nor do they always trace the sites that will evolve into the most
massive clusters by $z=0$. This is only partly due to uncertainties in stellar
mass estimates and largely reflects the stochastic nature of mass accretion, as
discussed in previous work \citep[e.g.][see also Fontanot et al. in
  preparation]{DeLucia_and_Blaizot_2007,Trenti_etal_2008,Overzier_etal_2009,Angulo_etal_2012}.

\begin{acknowledgements}
 We acknowledge the use of INAF-OATs computational resources within the
 framework of the CHIPP project \citep{Taffoni_etal_2020}.  MH acknowledges
 funding from the Swiss National Science Foundation (SNSF) via a PRIMA grant
 PR00P2-193577 ‘From cosmic dawn to high noon: the role of BHs for young
 galaxies’.
\end{acknowledgements}

\begin{appendix} 
\onecolumn
  \section{Additional figures}
\label{sec.app1}

\begin{figure*}
\centering
\resizebox{18cm}{!}{\includegraphics{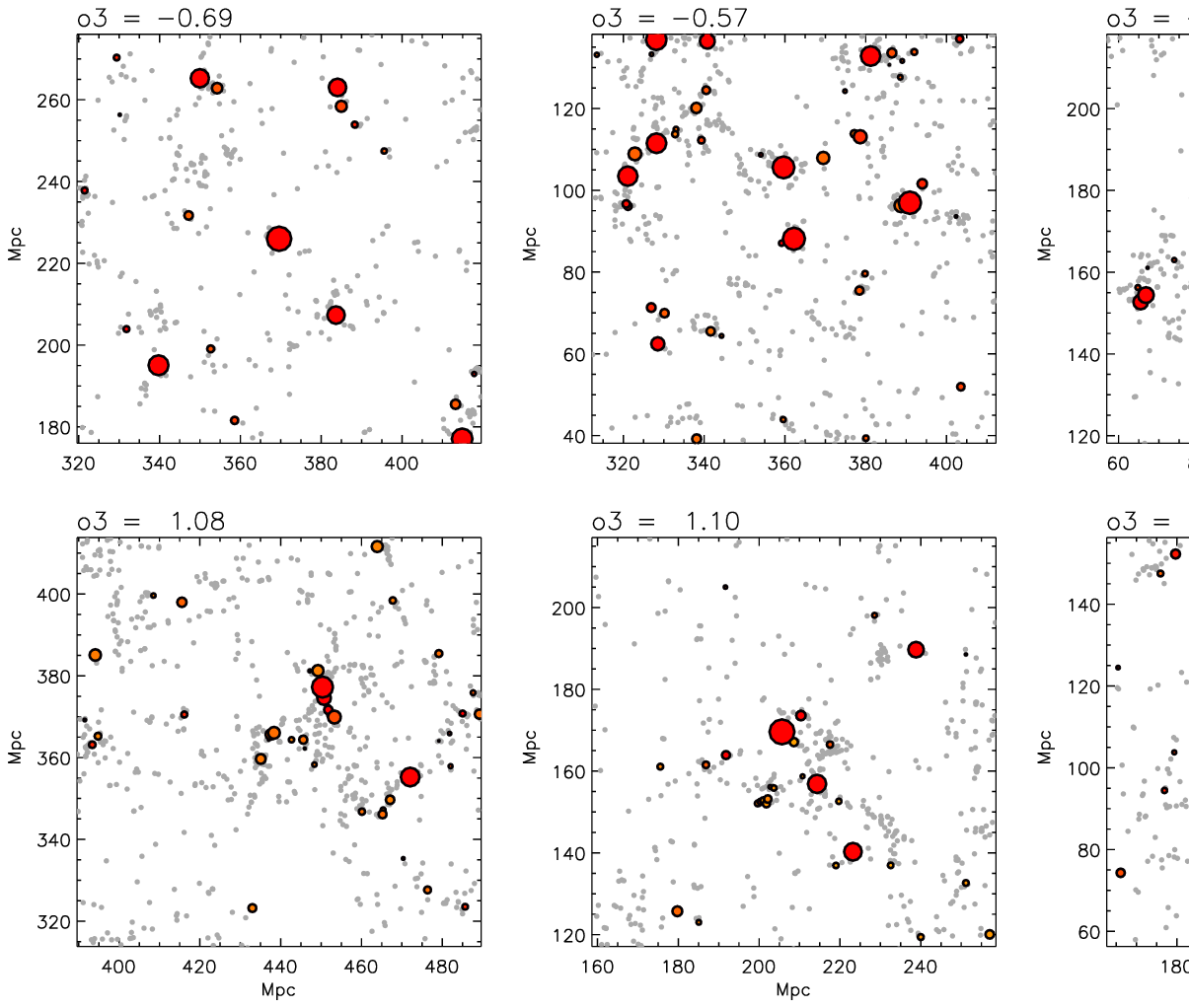}} 
\caption{As in Fig.~\ref{fig:xypos} but showing in color only galaxies with
  star formation rates lower than $\sim 5\,{\rm M}_{\odot}\,{\rm yr}^{-1}$.}
\label{fig:xyposlow}
\end{figure*}

\begin{figure*}
\centering
\resizebox{18cm}{!}{\includegraphics{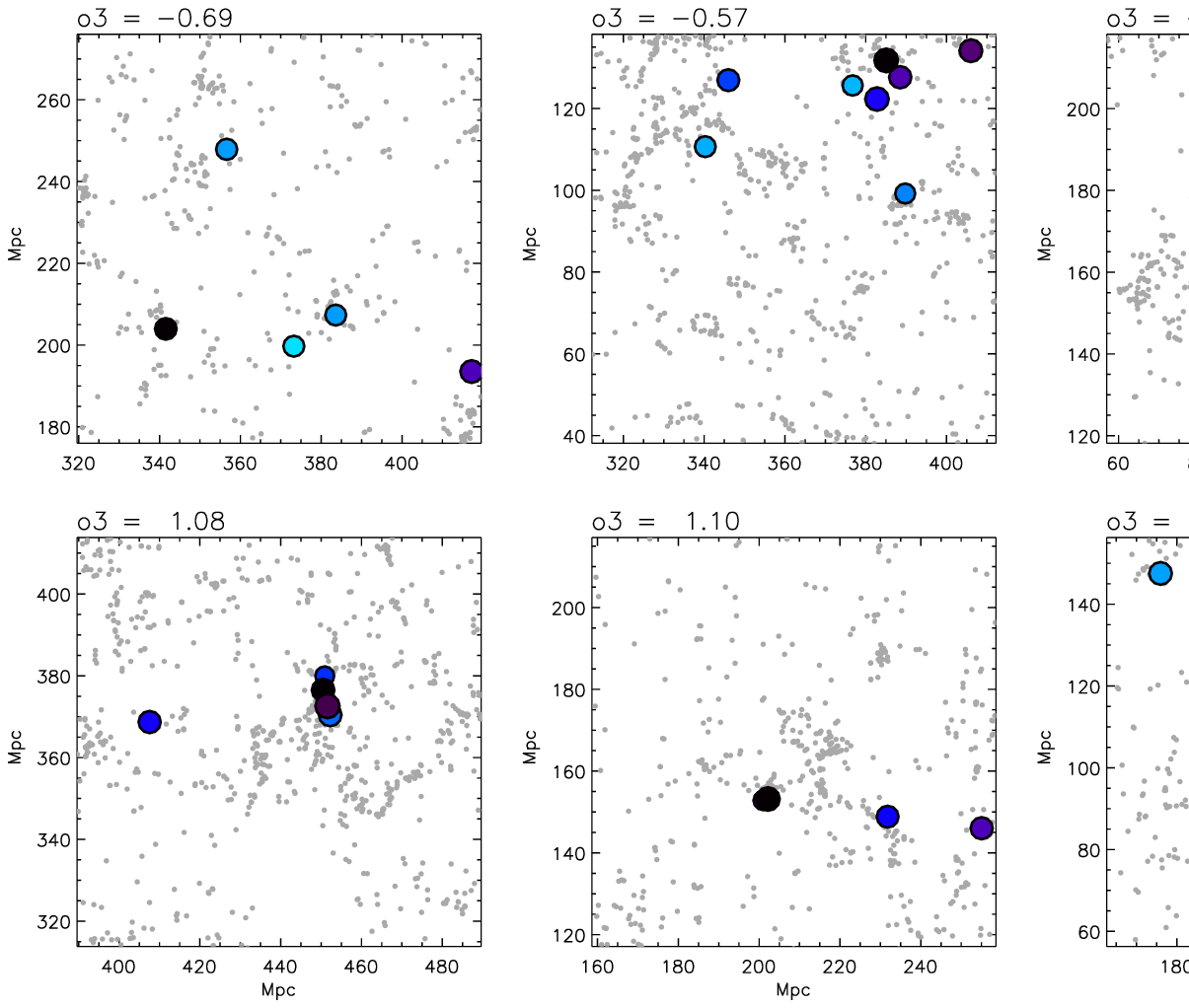}} 
\caption{As in Fig.~\ref{fig:xypos} but showing in color only galaxies with
  star formation rates larger than $\sim 80\,{\rm M}_{\odot}\,{\rm yr}^{-1}$.}
\label{fig:xyposhigh}
\end{figure*}

Figs.~\ref{fig:xyposlow} and \ref{fig:xyposhigh} show the projected positions
of galaxies with star formation rates lower than $\sim 5\,{\rm M}_{\odot}\,{\rm
  yr}^{-1}$ and larger than $\sim 80\,{\rm M}_{\odot}\,{\rm yr}^{-1}$,
respectively. The boxes correspond to those in Fig.~\ref{fig:xypos}, with gray
symbols showing all galaxies with stellar mass larger than $10^9\,{\rm
  M}_{\odot}$ independently of their star formation rate. A certain number of
galaxies forming stars at large rates can be found in all boxes considered. In
most cases these tend to avoid the central regions of the boxes but in a few
cases, due to projection effects and to the large projected volumes shown, they
appear to lie very close to the central regions.

\begin{figure*}
\centering
\resizebox{18cm}{!}{\includegraphics{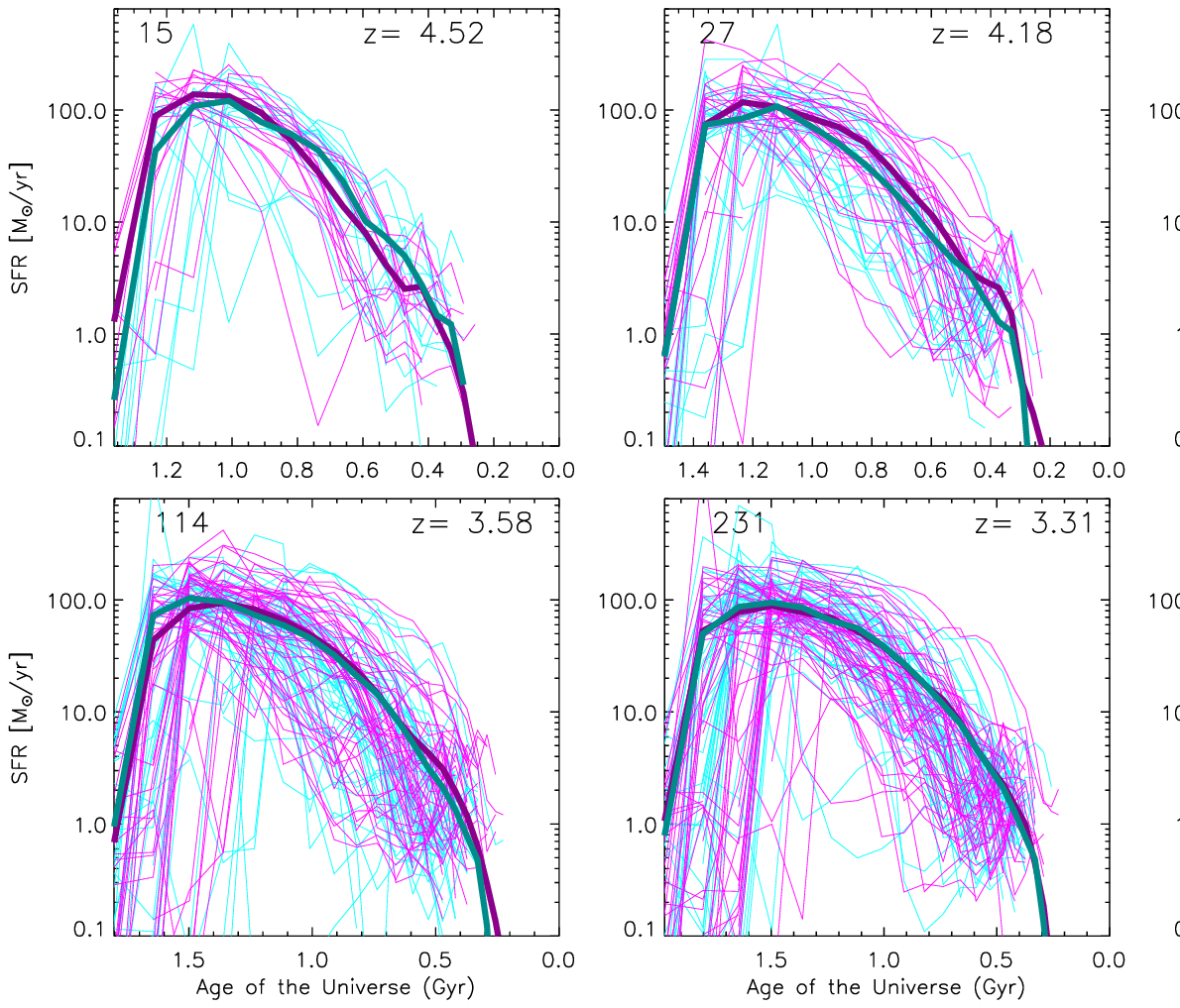}} 
\caption{The star formation histories of galaxies in the bottom 25th (cyan) and
  highest 75th (magenta) percentiles of the distribution of o$3$. Ticker lines
  show the median star formation histories. The numbers of galaxies in each
  sub-sample at each snapshot is given in the top left corner. A maximum of 50
  individual star formation histories in each subsample is shown to avoid
  overcrowding the panels.}
\label{fig:sfhextro3}
\end{figure*}

Fig.~\ref{fig:sfhextro3} is the equivalent of Fig.~\ref{fig:sfhextr} but
showing the average star formation histories of galaxies in our quiescent
sample in the 25th (dark cyan) and 75th (dark magenta) percentiles of the
distribution of o$3$. While there is a clear different (albeit with a very
large scatter) between the extremes considering the n$3$ distribution as
discussed in the main body of the paper, this difference is largely removed
when considering larger scale overdensities.

\begin{figure}
\centering \resizebox{18cm}{!}{\includegraphics{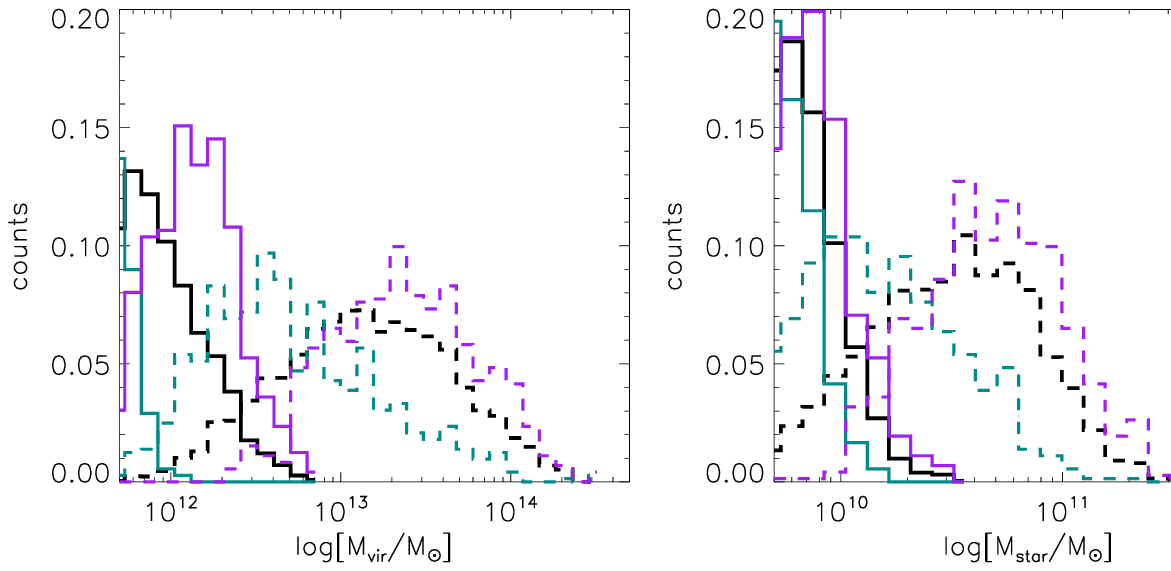}}
\caption{As in Fig.~\ref{fig:descmass}, but cyan and magenta histograms in the
  top and middle panel corresponds to galaxies in the extreme of the n$3$
  distribution. The bottom panel shows the correlation between the halo mass in
  which descendants of quiescent high-z galaxies reside at $z=0$ and n$3$ at
  the time of observation.}
\label{fig:descmassn3}
\end{figure}

Fig.~\ref{fig:descmassn3} is the equivalent of Fig.~\ref{fig:descmass} but for
galaxies in the extremes of the n$3$ distribution. The figure shows that n$3$
is a less powerful indicator of the $z=0$ halo mass: the scatter shown in the
bottom panel is significantly larger than that discussed in
Sec.~\ref{sec:future} when considering o$3$. However, the correlation is still
present so that galaxies in the extreme of the distribution still have
offsetted distributions of virial mass and stellar mass, as shown in the top
and middle panels.

\end{appendix}

\bibliographystyle{aa} 
\bibliography{envquiesc_delucia} 

\end{document}